\documentstyle[twocolumn,aps]{revtex}

\begin{document}

\twocolumn[\hsize\textwidth\columnwidth\hsize\csname
               @twocolumnfalse\endcsname

%\draft

%****************************    TITLE    *****************************

\title{Explicit construction of constrained instantons}

 %****************************     AUTHORS   ****************************

\author{Morten Nielsen$^\ast$,
N.K. Nielsen$^\dagger$.}
\address{Department of Physics, University of Southern Denmark,
Odense University, Denmark}

\date{Received 1 December, 1999 }
\bibliographystyle{unsrt}

\maketitle

%***************************     ABSTRACT    *****************************

\begin{abstract}
\noindent Instantons in massless theories do not carry over to massive
theories due to Derrick's
theorem. This theorem can, however, be circumvented,
if a constraint that restricts the scale of
the instanton is imposed on the theory.
Constrained instantons are considered in four dimensions in $\phi ^4$
theory and
$SU(2)$ Yang-Mills-Higgs theory. In each of these theories a calculational
sceme
is set up and solved in the lowest few orders in the mass parameter in
such a way that the need for a constraint is exhibited clearly.
Constrained instantons are shown to exist as finite action
solutions of the field equations with exponential fall off only for specific
constraints that are
unique in lowest order in the mass parameter in question.

\end{abstract}

\vskip2pc]

%\newpage

%***************************    TEXT       *******************************

%\section{Introduction}

%%%%%%%%%%%%%%%%%%%%%%%%%%%%%%%%%
\section{Introduction}
%%%%%%%%%%%%%%%%%%%%%%%%%%%%%%%%%%%%%%%%%
Instantons have been prominent tools for the computation of
nonperturbative effects in classically conformally invariant field theories
including
gauge theories since the pioneering
achievements of Belavin et al. \cite{Belavin} and
't Hooft \cite {'t Hooft}. In the presence of mass, including mass
generation because of spontaneous symmetry breaking, instantons
leading to a
finite action do not exist as a consequence of a generalization of
Derrick's theorem
\cite {Derrick}. However, as pointed out by Frishman and Yankielowicz
\cite {Frishman}
and Affleck
\cite {Affleck}, a finite action solution of the field theory in question can
be obtained if a constraint is imposed on the theory restricting the scale of
the instantons to be small compared to the inverse mass parameter.

Since then, constrained instantons have enjoyed considerable attention
\cite{Dine}, \cite {Espinosa}, \cite {Klinkhamer}, \cite{Wang}, \cite {Aoyama},
\cite{Dorokhov}.
However, little consideration has been given
to a systematic  explicit analytic construction of constrained instantons.

In the present paper, a detailed account is given of
the explicit construction of constrained instantons in the context of the two
models also considered in \cite {Affleck}, viz. ${\phi }^4$ theory with a
negative potential,
and $SU(2)$ Yang-Mills-Higgs theory. The latter example is especially
interesting because of its relevance for the standard model of electroweak
unification. The constructions in the two models are carried out
recursively in the mass parameters, following the pattern indicated in
\cite {Affleck} and in such a way that
the constrained instanton solutions at short distances do not contain
singularities spoiling the finiteness of the actions, while
their large-distance behavior is determined by the modified Bessel
function $K_1$, thus ensuring the exponential fall off familiar
from massive field theories.

\begin{center}
\line(1,0){240}
\end{center}

\vspace{1 mm}

\noindent $^\ast$Electronic address: morten@fysik.sdu.dk\\
$ ^\dagger$Electronic address: nkn@fysik.sdu.dk
\newpage
For $\phi ^4$ theory we find that the only way to achieve
this goal is by means
of a constraint cubic in the field or by having a similar constraint
through a source term in the
field
equation, while other constraints
only depending on the field
are ruled out
because they lead to singular behavior
of the constrained instanton solution at the origin.
For the Yang-Mills-Higgs theory  exponential fall off at infinity
can be obtained by adjustment of
integration constants, but a constraint is necessary for the Yang-Mills
field in order
to ensure absence of
singularities of the constrained
instanton at small distances that prevent the action from being
finite. No modification of the
Higgs field equation is necessary.

The important point about the analysis of the present paper is that the
accomplishment of a good constraint is twofold: It should
\begin{enumerate}
\item  restrict the scale
parameter of the instanton
solution, and
\item  ensure that the instanton solution leads to a finite action
\end{enumerate}
and 1. is a necessary but by no means sufficient condition for 2. In fact,
as we shall demonstrate, most constraints that ensure 1. lead to constrained
instanton solutions that are singular at the origin in such a way that 2.
is violated.
In lowest order in the mass parameter it is found that the form of the
constraint is uniquely fixed, whereas there is considerable freedom to
choose the constraint in higher orders.

The paper is organized in the following way: In Sec. \ref{phifour}
the scalar
$\phi^4$ theory is considered. Lowest-order corrections to the
instanton solutions due to a mass
parameter and constraint terms are calculated explicitly in
Sec. \ref{higher} and \ref{scacon}, and in these sections also a
leading-term analysis is carried out to all orders in the mass parameter,
showing that the constrained instanton solution leads to a finite action
and has exponential fall off at large distances. This argument
is completed in Sec. \ref{finac} and extended to subleading terms in
Sec. \ref{leadvssub}. In Sec. \ref{YMH} a corresponding analysis is
carried out for the Yang-Mills-Higgs theory. In Sec. \ref{finiteness}
conditions for finiteness of the action are obtained, and in Sec.
\ref{iteration} explicit
solutions of the field equations are found by iteration up to
fourth order in the mass parameter. In Sec. \ref{limit} a leading order
analysis is carried out on similar lines as in the scalar case. In Secs.
\ref{modeq}, \ref{constr} and \ref{lesub} a constraint leading
to the desired properties of the solution is constructed. In the course
of this construction also the subleading terms are considered. Finally
the connection to the 't Hooft integral measure is established in Sec.
\ref{thooft}.

%%%%%%%%%%%%%%%%%%%%%%%%%%%%%%%%%
\section{Scalar $\phi^4$ theory} \label{phifour}
%%%%%%%%%%%%%%%%%%%%%%%%%%%%%%%%%%%%%%%%%
In this section we consider an Euclidean $\phi^4$ theory with negative
$\phi ^4$ term. The Lagrangian is
\begin{equation}
L=-\frac{1}{g}\left(\frac{1}{2}(\partial _\mu \phi )^2
  +\frac12m^2\phi^2-\frac{1}{4!}\phi ^4\right).
\end{equation}
In the massless case the field equation has an instanton solution $\phi _0$:
\begin{equation}
\partial^2\phi _0+\frac16\phi _0^3=0,\hspace{2mm}
  \phi_0=\frac{4\sqrt{3}\rho}{\rho^2+x^2}.
  \label{zeroinst}
\end{equation}
Here and elsewhere $\rho $ is a scale parameter characterizing the
instanton solution.
The equation governing small deviations $\delta\phi$ from this solution
has the zero mode
$
\frac{\partial}{\partial\rho}\phi_0.
$
%%%%%%%%%%%%%%%%%%%%%%%%%%%%%%%%%%%%%%%%%%%
\subsection{Mass corrections}\label{higher}
%%%%%%%%%%%%%%%%%%%%%%%%%%%%%%%%%%%%%%
Take the mass $m$ to be small but non-zero.
This is expected to give rise to a small deviation $\phi_1$ from the massless
solution. It is convenient to introduce the variable $t=\frac{\rho^2}{x^2}$ in
terms of which the equation obeyed by $\phi _1$ is
\begin{equation}
\left(\frac{d^2}{dt^2}+\frac{6}{t(1+t)^2}\right)\phi_1=\sqrt{3}\rho m^2
  \frac{1}{t^2(1+t)}. \label{fe1}
\end{equation}
The operator on the left-hand side has a zero
mode $\frac{t(1-t)}{(1+t)^2}$
corresponding to $\frac{\partial}{\partial\rho}\phi_0$.
Introducing the Spence function \cite{Mitchell}
\cite {Hofreiter} by
\footnote
{The litterature contains several definitions of
the Spence function, differing mutually by signs and additive constants.
We have found the definition of (\ref{Spence}) most convenient. It leads to
the following identity:
$$
\Phi (t)+\Phi (\frac{1}{t})-\frac{1}{2}\log ^2t=\frac{\pi ^2}{6}.
$$}
\begin{equation}
\Phi(x)=\int_0^x\frac{du}{u}\log(1+u)
\label{Spence}
\end{equation}
a solution of (\ref{fe1}) is
\begin{eqnarray}
&&\phi_1=
  \sqrt{3}\rho m^2\left[
  \left(1+t-\frac{12t}{1+t}\right)\log\frac{1+t}{t} \right .
  \nonumber \\
  &&\left . +6\frac{t(1-t)}{(1+t)^2}\Phi\left(\frac{1}{t}\right)
  -\frac{12}{1+t}+9\right].
\label{deltaen}
\end{eqnarray}
To this solution may be added
a term proportional to the zero mode
$
\frac{t(1-t)}{(1+t)^2}.
$

In the limit $t\rightarrow \infty$ ($x\rightarrow 0$) the solution behaves as
a constant
\begin{equation}
\phi _1\simeq 10\sqrt{3}\rho m^2
\label{nulretning}
\end{equation}
which may be modified by a finite amount by adding to
the solution a multiple of the zero mode. In the opposite limit,
$t\rightarrow 0$ ($x\rightarrow \infty$), the outcome is
\begin{equation}
\phi _1\simeq \sqrt{3}\rho m^2\log \frac{1}{t}-3\sqrt{3}\rho m^2
\label{oneinst}
\end{equation}
that is unaffected by the zero mode.

Next consider higher order mass corrections by iteration of
the equation
\begin{equation}
(\partial^2-m^2)\phi=-\frac{1}{6}\phi ^3
\label{phi}
\end{equation}
in the mass parameter $m^2$. We are mainly concerned with the asymptotic
behavior in the regimes $x\rightarrow 0$ and $x\rightarrow \infty$. At
$x\rightarrow 0$ the mass corrections must be finite while at
$x\rightarrow \infty$ the leading mass corrections should sum
to
\begin{equation}
4\sqrt{3}\rho \frac{m}{x}K_1(m x),\label{ledende}\label{Bess}
\end{equation}
with $K_1$ a modified Bessel function, thus ensuring exponential
fall off of the instanton solution for large
$x$ according to (\ref{ask}). The factor in front is found by
comparison with the massless instanton solution (\ref{zeroinst}).
Exponential fall off of the
subleading terms also has to be achieved somehow. This problem
will be considered in Sec. \ref{leadvssub}.

Writing the solution of (\ref{phi}) as a power series in $m^2$ with
the term proportional to $m^{2n}$ denoted $\phi _n$, we observe from
(\ref{zeroinst}) the following behavior of $\phi _0$ at
large $x$:
\begin{equation}
\phi _0\simeq \frac{4\sqrt{3}\rho}{x^2},
\label{nulphi}
\end{equation}
while $\phi _1$ in this limit is given by
(\ref{oneinst}).
Thus it stands to reason that the leading terms to power $m^{2n}$
are proportional to $m^{2n}(x^{2})^{n-1}$
(for power-counting purposes logarithmic factors of
$x$ and $m $ can be disregarded,
as will be clear in the course of our argument).
 That this is indeed the case is proven  by
induction by means of (\ref{phi}), which is equivalent to:
\begin{equation}
\partial^2\phi_n-m^2\phi _{n-1}=-\frac{1}{6}\sum _{\nu_1+\nu _2+\nu _3=n}
\phi _{\nu _1}\phi _{\nu _2}\phi _{\nu _3}.
\label{phin}
\end{equation}
Assuming
\begin{equation}
\phi _i\propto m^{2i}(x^2)^{i-1}, \hspace{1 mm}i<n
\label{propto}
\end{equation}
we find that the term on the right-hand side involving $\phi _n$
is negligible compared to the first term on the left-hand side, whereas the
terms not involving $\phi _n$ are dominated by the second term on the
left-hand side.
Thus, to leading order the right-hand side of (\ref{phi})
can be safely neglected. It then follows that if (\ref{propto}) is valid to
order
$n-1$ then it also holds to order $n$, and hence to all orders by induction.

Eq.(\ref{phi}) with the nonlinear term on the right-hand side
disregarded is the
Klein-Gordon equation in four-dimensional Euclidean space. Consequently
the solution
is a linear combination of
$\frac{m}{x }I_1(mx)$ and
$\frac{m}{x }K_1(mx)$, with $I_1$ and $K_1$ modified Bessel
functions of the first and the second kind, respectively
(see appendix \ref{Besselapp}).  By comparison of
(\ref{sumi}) and
(\ref{sumk}), which for small values of $mx$ implies:
\begin{eqnarray}
&&\frac{m }{x}K_1(m x)=\frac{1}{x^2}+\frac{m^2}{4}
\left (\log\frac{m ^2x^2}{4}+2\gamma -1\right )+O(x^2),
\nonumber \\
&&\label{smallm}
\end{eqnarray}
with (\ref{oneinst}) and (\ref{nulphi}) we learn:
\begin{eqnarray}
&&\phi\simeq 2\sqrt{3}\rho \left (2\frac{m}{x}K_1(mx)\right .
\nonumber \\&&
\left. -\left (\log \frac{m^2\rho ^2}{4}+2\gamma+2\right )\frac{m}{ x}
I_1(mx)\right ).
\label{leading}
\end{eqnarray}
From (\ref{asi}) it is seen that the last term of (\ref{leading}) has
exponential
growth
at $x \rightarrow \infty$.
Thus this term prevents our solution from being a
finite action solution. This problem is expected according to the analysis of
\cite{Affleck} and is solved by imposing a constraint on the solution.

At $x\rightarrow 0$ it was found in (\ref{zeroinst}) and (\ref{nulretning})
that
$\phi _0$
and $\phi _1$ are regular. It  follows from (\ref{phi}) that all higher order
terms
are regular in this limit order by order in $m^2$. To see this we rewrite
the differential equation determining $\phi _n$ in terms of the variable
$t=\frac{\rho ^2}{x^2}$, obtaining an asymptotic equation of the form
\begin{equation}
(\frac{d^2}{dt^2}+\frac{6}{t(1+t)^2})H(t)=t^{-k},\hspace{0.1 mm}k\geq 3.
\end{equation}
The solution of this equation is
\begin{equation}
H(t)={\rm const}\times \frac{t(1-t)}{(1+t)^2}+\frac{1}{k(k-1)}t^{2-k}+O(t^{1-k})
\end{equation}
where the first term on the right-hand side originates from the zero-mode.
This solution, and consequently $\phi _n$, approaches a constant for
$t\rightarrow \infty$. In this way it follows by induction that the solution
of (\ref{phi}) is finite for
$x\rightarrow 0$ to all orders in $m^2$.

%%%%%%%%%%%%%%%%%%%%%%%%%%%%%%%%%%%%%%%%%%%
\subsection{Constraint corrections} \label{scacon}
%%%%%%%%%%%%%%%%%%%%%%%%%%%%%%%%%%%%%%
 Eq. (\ref{phi})
should now
be modified in such a way that the $I_1$-term in (\ref{leading}) is
eliminated while the regular behavior at $x\simeq 0$ is kept. According to
the prescription of  \cite{Affleck} this should be achieved by
introducing a constraint.

With a constraint
\begin{equation}
\int d^4x\phi ^{n}(x)=c\rho^{4-n}
\label{constance}
\end{equation}
with $n=3$ or $n\geq 5$ a positive integer,
the first order equation corresponding to (\ref{fe1}) is
\begin{equation}
\left(\partial^2+\frac{1}{2}\phi_0^2\right)\delta\phi _1=n
\bar{\sigma }(\phi_0)^{n-1}.
\label{qons}
\end{equation}
Here $c$ and $\bar{\sigma }$ are constants. The case $n=4$ is excluded (the
constraint in
this case does not break scale invariance, and this is a nessesary condition
for a finite action instanton solution
to exist).

For $n=3$ the solution is trivial:
\begin{equation}
\delta \phi _1=6\bar{\sigma }
\label{nligtre}
\end{equation}
This solution can be used to modify (\ref{oneinst})
in such a way that the unwanted $I_1$-term of (\ref{leading}) disappears, so in
this case one indeed obtains a
constrained instanton with an exponential fall off at large distances. The
details are given
below in Sec. \ref{finac}.

Introducing the variable $u=\frac{t}{1+t}$ and defining
\begin{equation}
F=\frac{1}{n\bar{\sigma}}\frac{4}{\rho^2}\left(\frac{\rho}{4\sqrt{3}}
\right)^{n-1}
\delta \phi _1\label{F}
\end{equation}
one converts (\ref{qons}) into an inhomogeneous hypergeometric equation
\begin{equation}
\left(u(1-u)\frac{d^2}{du^2}-2u\frac{d}{du}+6\right)F=u^{n-3}\;,\;\;n\geq 3.
\label{hyp}
\end{equation}
For completeness the solution is determined  also in the case $n=4$:
\begin{equation}
F=\frac{1}{4}u,
\end{equation}
 and for $n\geq 5$ the solution is:
\begin{eqnarray}
   &&F=\frac{u^{n-2}}{(n-2)(n-3)}
   \nonumber \\&&
   +\frac{n-4}{n(n-1)(n-2)}
   \sum_{i=n-2}^\infty
  \frac{(i+3)(i+2)}{i(i-1)}u^{i+1}.
\label{novertre}
\end{eqnarray}
To these solutions may be added an arbitrary multiple of the zero-mode
$\frac{t(1-t)}{(1+t)^2}=u(1-2u)$.

In the case $n\geq 5$ the terms in the infinite series tend from above to a
geometric series for large $i$, and
hence the expression diverges in the limit $u\rightarrow 1$, i. e.
$x\rightarrow 0$:
\begin{eqnarray}
\lim_{u \rightarrow 1}(1-u)F(u)=\frac{n-4}{n(n-1)(n-2)}.
\end{eqnarray}
 In consequence, we have found the following behavior of $\delta \phi _1$
in 
 the case
$n\geq 5$ :
\begin{eqnarray}
\delta \phi _1(x)&\simeq & 0\hspace{1 mm}{\rm for}\hspace{1
mm}x\rightarrow \infty,  \label{Fresult2}\\
\delta \phi _1(x)&\simeq &
\frac{\sigma \rho^4}{4}\left(\frac{\rho}{4\sqrt{3}}\right)^{1-n}
\frac{n-4}{(n-1)(n-2)}\frac{1}{x^2}
\hspace{1 mm}{\rm for}\hspace{1 mm}x\rightarrow 0.
\nonumber
\end{eqnarray}
In this case the constraint corrections lead to a singular
behavior of the instanton at $x\rightarrow 0$, invalidating
Affleck's equation (2.7) in \cite{Affleck}. It is instructive to see this
in detail.

In order to compare (\ref{Fresult2}) with eq. (2.7) in Affleck's paper
we use the following identity :
\begin{eqnarray}
&&\int d^4x\left(\frac{\partial \phi_0}{\partial\rho }\partial^2
\delta \phi _1-\delta \phi _1\partial^2
\frac{\partial \phi_0}{\partial\rho }\right)
\nonumber \\&&
=\int d^4x\partial _\mu \left(\frac{\partial \phi_0}{\partial\rho }
\stackrel{\leftrightarrow}{\partial } _{\mu }\delta \phi _1\right)
=\sigma \frac{\partial}{\partial\rho }
\int d^4x\left(\phi_0\right)^n.
\label{Gauss}
\end{eqnarray}
To obtain the last version of (\ref{Gauss}) we used that
$\frac{\partial \phi_0}{\partial\rho }$ is a zero mode, as well as (\ref{qons}).

The last version of (\ref{Gauss}) is nonvanishing. Rewriting the
middle version
by means of Gauss' theorem and assuming that there is no contribution from
a surface near
the origin one would conclude, following Affleck,  that $\delta \phi _1$
goes as a
nonvanishing constant
 for $x \rightarrow \infty $.

However, it follows from (\ref{Fresult2}) that a
contribution actually arises from a surface near the origin. This means that
$\delta \phi _1$
need not go as a constant for $x \rightarrow \infty $, and indeed
it vanishes in this limit according to (\ref{Fresult2}).

In the preceding paragraps we only considered constraints that are monomials in
the field $\phi $. For more generalized constraints the equation (\ref{hyp})
is replaced by
\begin{equation}
\left(u(1-u)\frac{d^2}{du^2}-2u\frac{d}{du}+6\right)F(u)=g(u)
\label{ghyp}
\end{equation}
where $g(u)$ is some function. If $g(u)$ can be expressed as a power series
it follows from the analysis of the preceding section that the solution
is singular for $x \rightarrow 0$
unless $g(u)$ only contains terms linear in $u$ or constant.

One example is the constraint\cite{Affleck}
$$
\int d^4x(\partial_\mu\phi\partial_\mu\phi)^{n}=c\rho^{4-4n}
$$
which leads to the following inhomogeneous hypergeometric equation
\begin{eqnarray}
&&\left(u(1-u)\frac{d^2}{du^2}-2u\frac{d}{du}+6\right)F(u)
\nonumber \\&&=-2(1+n)u^{3n-2}
  (1-u)^{n-1}
  \nonumber \\&&
  +6(n-1)u^{3n-3} (1-u)^{n}\;,\;\;n\geq 2 \label{flign}
\end{eqnarray}
replacing (\ref{hyp}).
Here the
right-hand side can be written as a power series in $u$
with cubic and higher-order terms, and the solution is consequently
singular for $x\rightarrow 0$.
This situation is similar to what was encountered for the constraint
(\ref{constance}) with $n \geq 5$.
Thus all these constraints must be rejected.

Another type of constraint, suggested by Frishman and Yankielowicz
\cite{Frishman}, corresponds to having an equation of the form
\begin{equation}
\left(u(1-u)\frac{d^2}{du^2}-2u\frac{d}{du}+6\right)F=\delta
(u-u_0)\;,\;\;n\geq 3
\label{hhyp}
\end{equation}
with $ u_0$ a constant between 0 and 1 and $\delta (u-u_0)$ the Dirac
$\delta $-function.
In this case one also expects $F(u)$ to be singular at $u=1$. In order
to prove this one can smear the constraint with a test function $g(u_0)$
and obtains then the situation considered previously.

%%%%%%%%%%%%%%%%%%%%%%%%%%%%%%%%%%%%%%%%%%%%%%%%%
\subsection{Construction of finite action constrained instanton}
\label{finac}
%%%%%%%%%%%%%%%%%%%%%%%%%%%%%%%%%%%%%%%%%%%%%%%%%%

In Sec. \ref{higher} it was found that the
leading terms of the solution sum to the result given in (\ref{leading})
where the last term, having exponential
growth at $x \rightarrow \infty$,  prevents the solution from being a
finite action solution and should be eliminated by means of a constraint.

For a constraint represented by a term in the action of the form
\begin{equation}
\sigma \left(\int d^4x\phi ^3(x)-c \rho \right),
\label{mortensaften}
\end{equation}
the constraint
 can be used to modify (\ref{oneinst}) in
such a way that the unwanted $I_1$-term of (\ref{leading}) disappears, so in
this case one indeed obtains a
constrained instanton with an exponential fall off at large distances.
Comparing (\ref{oneinst}) to (\ref{smallm})
we see from (\ref{nligtre}) that
if we fix the Lagrange multiplier according to
$\sigma =\bar{\sigma}$ with
\begin{equation}
\bar{\sigma}=\frac{\sqrt{3}}{6}\rho m^2\left (\log\frac{m^2\rho ^2}{4}
+2\gamma+2\right )
\label{segma}
\end{equation}
we have
\begin{equation}
\phi_1+\delta\phi_1\simeq \sqrt{3}\rho m^2
  \left(\log\frac{m^2x^2}{4}+2\gamma-1\right).
\end{equation}
This is exactly $4\sqrt{3}\rho\frac{m}{x}K_1(mx)$ to this order.

With (\ref{mortensaften}) added to the action and the Lagrange multiplier
$\sigma $ taking the value (\ref{segma}) one now obtains
$4\sqrt{3}\rho\frac{m}{x}K_1(mx)$ by summation of the leading terms
to all orders in the mass parameter. This
follows from:
\begin{itemize}
\item the analysis of Sec. \ref{higher} is unaffected by an extra term
proportional to $m^2\phi ^2$ on the right-hand side of the field equation;
the leading terms at
large $x$ still obey the Klein-Gordon equation.
\item $\frac{m}{x}I_1(mx)$ only contains a constant term in lowest order
(order $m^2$); thus if the constant is removed and the lowest-order term of
$\frac{m}{x}I_1(mx)$ thus is absent then all the higher-order terms must
also be absent.
\end{itemize}
The conclusion is that for a constraint (\ref{mortensaften})
 with
$\sigma $
given by (\ref{segma}) the sum of the leading terms has exponential fall off at
large distances.   The analysis of Sec. \ref{higher} on the finiteness of
the solution at small distances
is easily seen to be unaffected by the constraint.

%%%%%%%%%%%%%%%%%%%%%%%%%%%%%%%%%%%%%%%%%%%
\subsection{Leading vs. subleading terms} \label{leadvssub}
%%%%%%%%%%%%%%%%%%%%%%%%%%%%%%%%%%%%%%

Eq. (\ref{leading}) valid for $x\rightarrow\infty$ was found in a
leading term approximation, where only leading powers in $x^2$ were kept.
It should be checked that these leading terms are
still leading after summation. As was demonstrated in Sec.
\ref {finac},  the
$I_1$-term is removed by the constraint (\ref{mortensaften}).
The leading terms by themselves grow
faster the higher the order, but they conspire to a sum with exponential
fall off.
It is thus not  a priori clear that the nonleading terms are still
nonleading after summation. As we shall demonstrate, the sum of the nonleading
terms also have exponential fall off at large distances.

The field equation
\begin{equation}
(\partial^2-m^2)\phi+\frac{1}{6}\phi^3-3\sigma \phi ^2=0
\end{equation}
has in leading order at large distances the approximate solution
(\ref{ledende}), denoted $\phi^{(1)}$,
that is proportional to a massive scalar propagator
and is of first order in $\rho $ (hence the bracketed superscript).
The nextleading term $\phi^{(3)}$
in this
approximation scheme is a solution of the equation
\begin{eqnarray}
&&(\partial^2-m^2)\phi^{(3)}=-\frac{1}{6}(\phi^{(1)})^3 +6\sigma\frac{1}{2}
(\phi^{(1)})^2. \label{next}
\end{eqnarray}

The leading terms are of the form $m^{2n}(x^2)^{n-1}$, $n\geq 0$ and
possibly with
a logarithmic factor. The nextleading terms are correspondingly of the
form
$m^{2n}(x^2)^{n-2}$, $n\geq 0$ and again possibly with
a logarithmic factor. By inspection of (\ref{next}) it is seen that the
terms on the right-hand side are of
the form
$m^{2n}(x^2)^{n-3}$, so $\phi ^{(3)}$ is
indeed the sum of
the nextleading terms.

It is seen from (\ref{next}) that one can write $\phi^{(3)}$ as a
convolution
integral involving only massive propagators. Thus
$\phi^{(3)}$ also falls off exponentially at large distances. Continuing
this
approximation scheme one finds equations similar to
(\ref{next}) with the Klein-Gordon operator operating on
$\phi^{(2n+1)}$ on the left-hand side and an expression involving
previously found
$\phi^{(1)}$, $\phi^{(3)}$, $\cdots$ , $\phi^{(2n-1)}$ on the right-hand
side.
Hence $\phi^{(2n+1)}$ can be expressed as a convolution
integral involving only massive propagators as well
and exponential fall off
at infinity is ensured to each order.

In Sec. \ref{finac} a finite action solution was obtained by having
the constraint (\ref{mortensaften}). Having instead a
constraint
\begin{equation}
\sigma \left(\int d^4x3\phi _0 ^2(x)\phi (x)-c \rho \right)
\label{martinsaften}
\end{equation} clearly allows
the same conclusion, but only to leading order. The extra term in the
field equation is now a source
term instead of an expression quadratic in the field. In the equation
corresponding to (\ref{next}) this means that
the constraint induced term contains massless propagators. This in turn
means that
$\phi^{(3)}$ has a fall off at infinity according to a power law.

It is clearly desirable that the subleading terms have exponential
fall off like the leading terms. This can be obtained also
with a constraint leading to a source term in the field equation.
The reason is that the constant $\sigma $ in front of the
source term is of second order in the mass. It is
thus possible to use a constraint
$\sigma \int d^4x\hat{\phi } ^2(x)\phi (x)$, with
$$
\hat{\phi }=\phi _0+\cdots
$$
and where the higher-order terms are adjusted order by order such that
$\hat{\phi}=\phi$ with $\phi $ the solution obtained by
means of the constraint (\ref{mortensaften}). In this way a
source term constraint gives the same constrained instanton as a term
corresponding to
an extra term in the Lagrangian that is cubic in the scalar field.

Constraints corresponding to a source term in the field equation
were originally suggested by Wang \cite{Wang}.

%%%%%%%%%%%%%%%%%%%%%%%%%%%%%%%%%%%%%%%%%%%%%%%%%
\section{Yang-Mills-Higgs instanton} \label{YMH}
%%%%%%%%%%%%%%%%%%%%%%%%%%%%%%%%%%%%%%%%%%%%%%%%%%
An analysis of the SU(2) Yang-Mills-Higgs theory similar to that of the
$\phi ^4$
theory is carried out in this
section.
The Euclidean Lagrangian is
\begin{equation}
L=-\frac{1}{g^2}\left[\frac{1}{4}G_{\mu\nu}^aG_{\mu\nu}^a+\kappa
  \left((D_\mu\phi)^\dagger D_\mu\phi + \frac{1}{4}(\phi^\dagger\phi-\mu^2)^2
  \right)\right] \label{lagr}
\end{equation}
where
\begin{eqnarray}
G_{\mu\nu}^a&=&\partial_\mu A_\nu^a-\partial_\nu A_\mu^a
  +\epsilon^{abc}A_\mu^bA_\nu^c, \\
D_\mu&=&\partial_\mu-i\frac{\tau }{2}^aA_\mu^a
\end{eqnarray}
with $\tau ^a$ the Pauli matrices, and $\kappa >0$. The Yang-Mills field
acquires by the Higgs mechanism the mass
\begin{equation}
m_{vec }=\sqrt{\frac{\kappa }{2}}\mu .
\label{mvec}
\end{equation}
For $\mu =0$ the Yang-Mills field equation has the instanton solution in
the singular gauge
\begin{equation}
A_{0\mu}^a=\frac{2\rho^2\bar{\eta}^{a}_{\mu\nu}x_\nu}
{x^2(x^2+\rho^2)} \label{masslessinstanton}
\end{equation}
 with $\bar{\eta}^{a}_{\mu\nu}$ defined in \cite {'t Hooft}. With the
fields in the form
\begin{equation}
A^a_{\mu }
=-\bar{\eta }^{a}_{\mu \nu }\partial _{\nu }\log \alpha ,\hspace{2mm}
\phi =\left(\begin{array}{c}
  0\\f \end{array}\right)\label{ansatz}
\end{equation}
where $\alpha $ and $f$ are real functions
(the prepotential $\alpha $ was introduced in \cite {Jackiw}), the field
equations reduce to
\begin{eqnarray}
\alpha\partial_\nu\partial^2\alpha-3\partial_\nu\alpha\partial^2\alpha=
  \frac{\kappa }{2}f^2\alpha\partial_\nu\alpha ,
\label{alphaeq}
\end{eqnarray}
\begin{eqnarray}
\alpha^2\partial^2f-\frac{3}{4}(\partial_\nu\alpha)^2f
  +\frac{1}{2}\alpha^2f\left(\mu^2-f^2\right)=0.
\label{feq}
\end{eqnarray}
These equations are solved order by order in $\mu $ in such a way
that the solution leads to a finite action.
It is assumed that terms in $\alpha$ are of even order, while terms in $f$
are of
odd order in $\mu $.
The exponential decay at infinity can be ensured by tuning
of the integration constants in the first few orders, while integrability
of the Lagrangian density
at the
origin requires modification of the above equations corresponding to a
constraint.

%%%%%%%%%%%%%%%%%%%%%%%%%%%%%%%%%%%%%%%%%%%%%%%%
\subsection{Conditions for finiteness of the action} \label{finiteness}
%%%%%%%%%%%%%%%%%%%%%%%%%%%%%%%%%%%%%%%%%%%%%%%%%%%%%

With the fields only depending on the norm $x=|x|$
and in terms of the parameter $t=\frac{\rho^2}{x^2}$
the Lagrangian (\ref{lagr}) is
\begin{eqnarray}
&&L=-\frac{1}{g^2}\left[\frac{3}{2}\left(a^2
  +\left(\frac{b\rho^2}{t}-a\right)^2\right)
  +\kappa\frac{4t^3}{\rho^2}\left(\frac{df}{dt}\right)^2 \right .
  \nonumber \\ && \left .
  +\kappa \frac{3t^3}{\rho ^2}f^2
    \left(\frac{1}{\alpha }\frac{d\alpha}{dt}\right)^2
  +\frac{\kappa }{4}\left(f^2-\mu ^2\right)^2\right] \label{lagra}
\end{eqnarray}
where
\begin{equation}
a=\frac{4}{\rho ^2}\left(t^3
\left(\frac{d\log \alpha}{dt}\right)^2-t^2\frac{d\log \alpha}{dt}\right),
\label{aequ}
\end{equation}
\begin{equation}
b=\frac{2ta}{\rho^2}-\frac{4t^4}{\rho ^4}\frac{1}{\alpha }
  \frac{d^2\alpha}{dt^2}.
\label{rel} \label{bequ}
\end{equation}
The Lagrangian is negative semi-definite since the square bracket in
(\ref{lagra}) can
be expressed as a sum of squares.
Thus the condition for a finite action is that each term in the sum
gives a finite contribution  separately.

$f$ and $\log \alpha $ should at large $x$-values decrease
exponentially. Writing
\begin{equation}
\alpha =\alpha _0+\alpha _2+\alpha _4+\cdots,
\end{equation}
and similarly
\begin{equation}
f=f_1+f_3+f_5+\cdots
\end{equation}
with the indices enumerating the power of the mass parameter, and
\begin{equation}
\alpha _0=1+t
\label{alfanul}
\end{equation}
one obtains the following  estimate at small $t$-values:
\begin{equation}
\log \alpha \simeq (t+\alpha _2+\alpha _4+\cdots)
  -\frac{1}{2}\left(t+\alpha _2+\alpha _4+\cdots\right)^2+\cdots .
\label{logalfa}
\end{equation}
To ensure the exponential decay at small $t$-values one should express the
first bracket in this expression by the modified Bessel function $K_1$. To
make $\log\alpha$ behave order by order as a power series expansion of
$\frac{m}{x}K_1(m x)$ with $m$ some mass parameter
we have to require absence of inverse powers of $t$ in $\alpha _2$, while
$\alpha _4$ only is allowed to have a $1/t$-term etc. Similar considerations
apply to $f$. Thus the leading terms  should in the
limit $t\rightarrow 0$ conspire to  modified Bessel funtions according to
\begin{equation}
\alpha=\alpha_0+\alpha_2+\alpha_4+\cdots\simeq 1+\frac{m_{vec}\rho^2}{x}
K_1(m_{vec}x). \label{an1}
\end{equation}
For $f$ the appropriate condition turns out to be
\begin{equation}
f=f_1+f_3+f_5+\cdots\simeq\mu-\frac{\rho^2\mu^2}{2x}K_1(\mu x),
\label{an2}
\end{equation}
as demonstrated explicitly in Sec. \ref{limit}.

To investigate the integrability at the origin of the
 Lagrangian density (\ref{lagra}) one first considers:
\begin{eqnarray}
a^2=a_0^2+2a_0a_2+a_2^2+2a_0a_4+\cdots
\end{eqnarray}
and the same for $\left(\frac{b\rho^2}{t}-a\right)^2$.
 From (\ref{alfanul}) follows that in
zeroth order:
\begin{equation}
a_0=b_0x^2-a_0=-\frac{4}{\rho ^2}\frac{t^2}{(1+t)^2}
\end{equation}
so $a_0^2$ and $\left(\frac{b_0\rho^2}{t}-a_0\right)^2$ are bounded for
$t\rightarrow \infty $.
Next
\begin{equation}
a_2=\frac{4}{\rho ^2}(\frac{2t}{1+t}-1)t^2\frac{d}{dt}\frac{\alpha _2}{1+t}
\end{equation}
from which it is deduced that to keep $a_0a_2$ integrable at the origin
 one can at most allow $\alpha_2$ to go as $t\log t$ at large $t$.
 If this condition is fulfilled then the
second square of (\ref{lagra}) is integrable at the origin to second order
as well.

In fourth order both squares contain two terms. If there is a term
$t\log t$
present in $\alpha_2$ then $a_2^2$ is not integrable at the origin;
it contains a term
quadratic in $t$. This must be cancelled by the other fourth order term in
$a^2$.
To determine $a_0a_4$
one finds from (\ref{aequ}):
\begin{eqnarray}
&&a_4=\frac{4}{\rho ^2}\left( t^3\left(\frac{d}{dt}
  \frac{\alpha _2}{1+t}\right)^2\right.
  \nonumber \\ &&\left.
  +(\frac{2t}{1+t}-1)t^2\left(\frac{d}{dt}\frac{\alpha _4}{1+t}
  -\frac{\alpha _2}{(1+t)^2}\frac{d\alpha _2}{dt}\right)\right)
\end{eqnarray}
To ensure a finite integral of $a^2$ to fourth order there
must be present
in $\alpha_4$ a term quadratic in $t$. Now examine the second square
of (\ref{lagra}) in a
similar
way. It is found that $\left(\frac{b_2\rho^2}{t}-a_2\right)^2$ is
integrable at
the origin,
whereas
\begin{equation}
\left(\frac{b_0\rho^2}{t}-a_0\right)\left(\frac{b_4\rho^2}{t}-a_4\right)
\end{equation}
is finite upon integration only if $\alpha_4$ diverges no faster than $t\log t$.
Thus we conclude that to have a finite action it is necessary to demand that
$\alpha_2$ diverges at most logarithmically for large values of the variable
$t$.
Continuing this analysis to higher
orders reveals that $\alpha_{2n}$ can at most diverge logarithmically
at large $t$ for
all $n>0$.

From the third square in (\ref{lagra}) it follows immediately that
$f$ can also at most grow as $\log t$ for $t\rightarrow\infty$.
This also ensures the integrability at the origin of the final two terms in
(\ref{lagra}), and thus of the entire Lagrangian.

%%%%%%%%%%%%%%%%%%%%%%%%%%%%
\subsection{Iteration}
\label{iteration}
%%%%%%%%%%%%%%%%%%%%%%%%%%
In this section the field equations are solved order by order in the
mass parameter $\mu $. In each order it is examined whether this
solution
has exponential fall off at infinity, as well as integrability of the
Lagrangian density
at the
origin, following the considerations of the previous section.
We start by rewriting (\ref{alphaeq}) and (\ref{feq}) in terms of the
parameter $t=\frac{\rho^2}{x^2}$:
\begin{eqnarray}
  \frac{d}{dt}\left(\alpha^{-3}t^3\frac{d^2\alpha}{dt^2}\right)
  =\frac{\kappa\rho^2}{8}
  f^2\alpha^{-3}\frac{d\alpha}{dt}, \label{li1}
\end{eqnarray}
\begin{eqnarray}
  \alpha^2\frac{d^2f}{dt^2}-\frac{3}{4}\left(\frac{d\alpha}{dt}\right)^2f
  =\frac{\alpha^2f\rho^2}{8t^3}(f^2-\mu^2).  \label{li2}
\end{eqnarray}
The equations can then in each step be solved by quadrature.
The integration constants arising this way are denoted $c_{i;j}$ where
the first subscript indicates the order
and the second subscript is an extra label.

%%%%%%%%%%%%%%%%%%%%%%%%%%%%
\subsubsection{Orders zero, one, two and three}
\label{the four lowest orders}
%%%%%%%%%%%%%%%%%%%%%%%%%%

To order zero the solution of (\ref{li1}) is (\ref{alfanul}),
 corresponding to
 the massless instanton in the singular gauge,
while to first order one obtains the following solution of (\ref{li2})
\begin{eqnarray}
f_1=\frac{c_{1;1}}{\sqrt{1+t}}+c_{1;2}(1+t)^{\frac{3}{2}}.
\label{fet}
\end{eqnarray}
It is necessary to choose $c_{1;2}=0$ in order to keep $f_1$ bounded for
$x\rightarrow 0$. A similar term will arise in each order and must
always be chosen equal to zero.
Eq. (\ref {fet}) reduces for $c_{1;1}=\mu$ to the isospin
$\frac{1}{2}$ zero mode \cite {'t Hooft}. It will be shown that this value of
$c_{1;1}$ is enforced by the boundary conditions.

In second order the solution of (\ref{li1}) is
\begin{eqnarray}
&& \alpha_2=
\frac{1}{2}\left(c_{2;1}-\frac{1}{3}\frac{\kappa\rho^2}{8}c_{1;1}^2\right)\frac
{1}{t}
  -3c_{2;1}\ln t+3c_{2;1}t\ln t
  \nonumber \\ &&
  +\frac{1}{2}c_{2;1}t^2+
(c_{2;2}-3c_{2;1})t+c_{2;3}.
\label{alfatofoerst}
\end{eqnarray}
 According to
the discussion  after (\ref{logalfa}) the first term must vanish, i.e.
\begin{equation}
c_{2;1}=\frac{1}{3}\frac{\kappa\rho^2}{8}c_{1;1}^2.
\end{equation}
The integration constant $c_{2;2}$ is taken equal $3c_{2;1}$
(a different choice of $c_{2;2}$ corresponds to a different
scale of $\rho $).
The terms $\frac{1}{2}c_{2;1}t^2$ and $3c_{2;1}t\ln t$ have to be eliminated
for a finite action solution according to the discussion after (\ref{logalfa}).
This can be accomplished by modifying the equation determining $\alpha _2$
according to:
\begin{eqnarray}
&&\frac{d}{dt}\left(\left(\frac{t}{1+t}\right )^3\frac{d^2\alpha_2}{dt^2}\right)
  +\frac{\kappa\rho^2c_{1;1}^2}{4}
\frac{t}{(1+t)^4}
 =\frac{\kappa\rho^2c_{1;1}^2}{8(1+t)^4}
\label{alfa2mod}
\end{eqnarray}
 resulting in
\begin{eqnarray}
\alpha_2=
  -\frac{\kappa\rho^2}{8}c_{1;1}^2\log t+c_{2;3}.
\end{eqnarray}
At small $t$-values $\log\alpha$ should vanish like
$\frac{m}{x}K_1(mx)$ with $K_1$ a modified Bessel function;
see (\ref{smallm}).
This is matched for
\begin{equation}
c_{2;3}=\frac{\kappa \rho ^2 c_{1;1}^2}{8}
(\log \frac{\kappa \rho^2c_{1;1}^2}{8}+2\gamma -1)
\end{equation}
that leads to the following form of $\alpha _2$:
\begin{equation}
\alpha _2= \frac{\kappa \rho ^2 c_{1;1}^2}{8}
(\log \frac{\kappa \rho ^2c_{1:1}^2}{8t}+2\gamma -1).
\label{alefato}
\end{equation}
 The mass parameter $m$ in the modified Bessel function is for
 $c_{1;1}=\mu $
actually equal to $m_{vec}$, the vector mass generated by the
Higgs field (\ref{lagr}).

The modification of (\ref{li1}) as displayed
in (\ref{alfa2mod}) is an indication of the necessity
of a constraint in the sense of \cite{Affleck}.
However, it should be emphasized that the modification is unique to second
order
in the mass parameter $\mu $. Any other modification will either cause
$\alpha _2$
to behave differently from the modified Bessel function $K_1$ at infinity,
 since it will modify the coefficient of the
$\log t$-term  or introduce more singular terms for $t\simeq 0$,  or it will
give rise to nonintegrable singularities of
the action density at the origin. This point is further elaborated upon in Sec.
\ref{modeq}.

In third order the solution of (\ref{li2}) is
\begin{eqnarray}
&&f_3=
  -\frac{\kappa \rho ^2  c_{1;1}^3}{16(1+t)^{\frac{3}{2}}}
  \left(\log \frac{\kappa \rho ^2c_{1;1}^2}{8t}+2\gamma-1\right)
  \nonumber \\ &&
  +\frac{\rho ^2c_{1;1}}{8\sqrt{1+t}}(\mu ^2-\frac{\kappa} {2}c_{1;1}^2)
  \left(\frac{3}{2}+t-(1+t)^2\log(1+\frac{1}{t})\right)
 \nonumber \\
  &&+\frac{\rho ^2c_{1;1}}{16\sqrt{1+t}}\left(c_{1;1}^2-\mu ^2\right)
  \left(\frac{(1+t)^2}{t}+2(1+t)+\frac{1}{2}
  \right .
  \nonumber \\ &&
  \left .
  -3(1+t)^2\log(1+\frac{1}{t})
  \right)
  - \frac{ c_{3;1}}{2\sqrt{1+t}}.
\label{tildeftre1} \label{vildeftre}
\end{eqnarray}
$f$ should at $x\rightarrow \infty $  behave  as
a modified Bessel
function $K_1$:
\begin{eqnarray}
f(x)-\mu\propto \frac{m}{x}K_1(m x)
\end{eqnarray}
with $m$ some mass parameter.
The small $m$-expansion of the Bessel function (\ref{smallm})
is used for the determinination of $c_{3;1}$ and $c_{1;1}$.
The mass parameter is also fixed this way.
Making a small-$t$ expansion of $f_1$ one finds
\begin{eqnarray}
f_1=c_{1;1}-\frac{c_{1;1}\rho^2}{2x^2}+\cdots .
\label{fat}
\end{eqnarray}
Thus, $f_3$ must for $t\rightarrow 0$ reduce to
\begin{equation}
f_3\simeq
-\frac{c_{1;1}\rho^2}{2}\frac{m^2}{4}\left(\log\frac{m ^2\rho ^2}{4t}+2\gamma-1
\right).
\end{equation}
From this expression it is first observed that no term proportional to
$\frac{1}{t}$ occurs. Comparing with (\ref{tildeftre1}) one
immediately concludes
\begin{equation}
c_{1;1}=\mu .
\end{equation}
(\ref{tildeftre1}) then reduces in
the small-$t$ limit to
\begin{eqnarray}
&&f_3
  \simeq
  -\frac{\kappa \rho ^2 \mu ^3 }{16}
  \left(\log \frac{\kappa \rho ^2\mu ^2}{8}+2\gamma-1\right)
  \nonumber \\ &&
  +\frac{3\rho ^2\mu ^3}{16}(1-\frac{\kappa }{2})+\frac{\rho ^2\mu ^3}{8}\log t
  -\frac{c_{3;1}}{2}.
\end{eqnarray}
From the term involving $\log t$ one sees that the mass parameter $m$ must be
identified with the Higgs mass $\mu $. Notice that this identification
is enforced by the boundary condition.
Also the following
value of the integration constant $c_{3;1}$ is found:
\begin{eqnarray}
&&c_{3;1}=\frac{\rho^2\mu^3}{4}(1-\frac{\kappa }{2})
\left(\log\frac{\rho ^2\mu ^2}{4}+2\gamma+\frac{1}{2}
  \right)\nonumber \\&&-\frac{\kappa \rho ^2 \mu ^3}{8}\log \frac{\kappa }{2}.
\end{eqnarray}
Hence $f_3$ is completely determined.

%%%%%%%%%%%%%%%%%%%%%%%%%%%%
\subsubsection{Order four} \label{fourth order}
%%%%%%%%%%%%%%%%%%%%%%%%%%

$\alpha _4$ is according to (\ref{li1}) a solution of:
\begin{eqnarray}
&&\frac{d}{dt}\left(\left(\frac{t}{1+t}\right )^3\frac{d^2\alpha_4}{dt^2}
  -3\alpha_2\frac{t^3}{(1+t)^4}\frac{d^2\alpha_2}{dt^2}\right)
 \nonumber \\ &&=\frac{\kappa\rho ^2\mu ^2}{8}\left(\frac{d\alpha_2}{dt}
    \frac{1}{(1+t)^4}-\frac{3\alpha_2}{(1+t)^5}
    +\frac{2f_3}{\mu (1+t)^\frac{7}{2}}\right)
 \label{alpha4}
\end{eqnarray}
whence by insertion of $\alpha _2$ and $f_3$ and performing the first
integration:
\begin{eqnarray}
&&\frac{d^2\alpha_4}{dt^2}=\frac{\kappa^2\rho ^4\mu ^4}{64}
  \left(\log\frac{\kappa \rho ^2\mu ^2}{8t}+2\gamma-1\right)
  \left(\frac{1}{t^3}+\frac{2}{t^2(1+t)}\right)
  \nonumber \\ &&
  -\frac{\kappa\rho ^4\mu ^4}{32}\left(1-\frac{\kappa}{2}\right)
  \left(\left(\frac{1+t}{t}\right )^2\log \frac{1+t}{t}
  -\frac{5}{6}\frac{1}{t^3}-\frac{3}{2}\frac{1}{t^2}-\frac{1}{t}\right)
  \nonumber \\ &&
  -\frac{\kappa \rho ^4\mu ^4}{96t^3}
  \left((\frac{\kappa}{2}-1)\left(\log\frac{\rho ^2\mu ^2}{4}
   +2\gamma-1\right)+\frac{\kappa}{2}\log\frac{\kappa }{2}\right)
   \nonumber \\ &&
  +c_{4;1}\left (\frac{1+t}{t}\right )^3.
\label{determ4}
\end{eqnarray}
Letting $t\rightarrow\infty$ and disregarding all terms which vanish as
$t^{-2}$ or faster results in
\begin{eqnarray}
\frac{d^2\alpha_4}{dt^2}&\simeq&
  c_{4;1}\left(\frac{3}{t}+1\right)
\label{c41}
\end{eqnarray}
that gives rise to the following terms in $\alpha _4$:
\begin{equation}
\frac{1}{2}c_{4;1}t^2+3c_{4;1}t\log t
\end{equation}
that grow too fast for $t\rightarrow \infty $ to allow a
finite action solution. However, we cannot take the integration constant
$c_{4;1}$ equal to zero; indeed we find below that it has to have
a nonzero value in order that
$\log \alpha $ behaves as a Bessel function for
$t\rightarrow 0$.

A similar problem was encountered for $\alpha _2$ where it led to the modified
differential equation (\ref{alfa2mod}).
Modifying the differential equation (\ref{determ4}) in the same
way means removing from its right-hand side the terms on the right-hand
side of (\ref{c41}).
Integrating twice the resulting equation one finds the solution
\begin{eqnarray}
&&\alpha_4=\frac{\kappa^2\rho ^4\mu ^4}{64}
  \left(\log\frac{\kappa\rho ^2\mu ^2}{8t}+2\gamma-1\right)\times
  \nonumber \\ && \times
  \left(\frac{1}{2t}+2(1+t)\log \frac{1+t}{t}\right)
  \nonumber \\ &&
  -\frac{\kappa^2 \rho ^4\mu ^4}{64}
  \left[\frac{3}{4t}
 -2\log t
  +2(1+t)\Phi\left (\frac{1}{t}\right )\right]
  \nonumber \\ &&
  +\frac{\kappa\rho ^4\mu ^4}{32}\left(1-\frac{\kappa}{2}\right)
  \left[\frac{1}{2}(1+t)(5-t)\log\frac{1+t}{t} \right .
  \nonumber \\ && \left .
  -(1-2t)\Phi\left (\frac{1}{t}\right )
  +\frac{t}{2}+\frac{5}{12t}
  \right] \nonumber
  \\  &&
  -\frac{\kappa \rho ^4 \mu ^4}{192t}
  \left((\frac{\kappa}{2}-1)\left(\log\frac{\rho ^2\mu ^2}{4}
   +2\gamma-1\right)+\frac{\kappa}{2}\log\frac{\kappa }{2}\right)
  \nonumber \\ &&
  +c_{4;1}\left(\frac{1}{2t}
  -3\log t\right)
  +c_{4;2}t+c_{4;3} \label{CC}
\end{eqnarray}
with $\Phi (x)$ the Spence function defined in (\ref{Spence}).

The asymptotic behavior at $t\rightarrow 0 $ of (\ref{CC}) must be equal to
terms of order
$m_{vec}^4$ in $\rho^2\frac{m_{vec}}{x}K_1(m_{vec}x)$:
\begin{eqnarray}
\frac{\kappa^2\rho ^4\mu ^4}{128}\left(\log\frac{\kappa \rho ^2\mu ^2}{8t}
  +2\gamma-\frac{5}{2}\right)
  \frac{1}{t}.
  \label{besselfire}
\end{eqnarray}
While the $\frac{\log t}{t}$ terms match immediately, the terms of form
$\frac{\rm constant}{t}$ have to be adjusted by means of the
integration
constant $c_{4;1}$ according to
\begin{eqnarray}
&&c_{4;1}=-\frac{\kappa \rho ^4\mu ^4}{96}(1-\frac{\kappa }{2})
\left(\log\frac{\rho ^2\mu ^2}{4}+2\gamma+\frac{3}{2}\right)
\nonumber \\&&
+\frac{\kappa ^2\rho ^4\mu ^4}{192}\log\frac{\kappa }{2}.
\label{cIVI}
\end{eqnarray}

For large $t$ the asymptotic form
of $\alpha _4$ is according to (\ref{CC}):
\begin{eqnarray}
\alpha_4&\simeq
&c_{4;2}t.
\end{eqnarray}
Here one should take
\begin{equation}
c_{4;2}=0
\end{equation}
in order to ensure acceptable behavior of $\alpha _4$ at
$t\rightarrow \infty $. The constant $c_{4;3}$ is arbitrary.

To summarize, we have found to order $\mu ^4$ that a finite
action solution
exists if (\ref{li1})
is modified to:
\begin{eqnarray}
  \frac{d}{dt}\left(\alpha^{-3}t^3\frac{d^2\alpha}{dt^2}\right)-
\frac{3\bar{\sigma }t}{2(1+t)^4}
  =\frac{\kappa\rho^2}{8}
  f^2\alpha^{-3}\frac{d\alpha}{dt}  \label{alphaeq1}
\end{eqnarray}
with
\begin{eqnarray}
&&\bar{\sigma}=-\frac{\kappa \rho ^2 \mu ^2}{6}
+\frac{\kappa \rho ^4\mu ^4}{24}(1-\frac{\kappa }{2})
\left(\log\frac{\rho ^2\mu ^2}{4}+2\gamma+\frac{3}{2}\right)
\nonumber \\ && -
\frac{\kappa ^2\rho ^4\mu ^4}{48}\log \frac{\kappa }{2}
  +O(\mu ^6),
\label{alphaeq2}
\end{eqnarray}
and here Bessel function behavior of the solution at large
distances has been obtained by a suitable choice of the integration
constants.

%%%%%%%%%%%%%%%%%%%%%%%%%%%%%%%%%
\subsection{Limit considerations} \label {limit}
%%%%%%%%%%%%%%%%%%%%%%%%%%%%%%%%
Next (\ref{li1}) and (\ref{li2})  are examined
in the limits where $t\rightarrow 0$ and $t\rightarrow \infty$ in
order to reach some conclusion which are valid to all orders in the mass,
in the same way as in Sec.  \ref{higher}.
In these limits the equations simplify sufficiently to allow a leading term
analysis.
%%%%%%%%%%%%%%%%%%%%%%%%%%%%%%%%%%%%%%
\subsubsection{$t\rightarrow 0$} \label{limit0}
%%%%%%%%%%%%%%%%%%%%%%%%%%%%%%%%%%%%%%%%
Here it is checked by induction that the leading
terms of $\alpha $
and $f$ in powers
of $t$ conspire to give the modified Bessel function $K_1$ according to
(\ref{an1}) and (\ref{an2}).
More specifically, it will be checked that these expressions agree with the
leading
terms in the field equations. $\alpha _0$ is given in (\ref {alfanul})
while  $f_1$ at large distances behaves
according to (\ref{fat}).
The induction hypothesis is
\begin{equation}
\alpha_n\propto t^{1-\frac{n}{2}},\;n > 0
\;,\;\;f_n\propto t^{\frac{3}{2}-\frac{n}{2}},\;
n > 1. \label{indu}
\end{equation}
This hypothesis is correct for $n=2,3,4$ according to
(\ref{alefato}) (with $c_{1;1}=\mu $), (\ref{vildeftre}) and (\ref{CC}).
We want to prove it by induction for $n>4$ and to show that the leading terms
sum to (\ref{an1}) and (\ref{an2}).

For $n\geq 2$ one of the leading terms includes a logarithmic
factor, but this makes no
difference in what follows since it does not affect the estimate of the
power behavior after differentiation.

Keeping only the leading terms for $t\rightarrow 0$ one gets from (\ref{li1})
to order $n$ in the mass parameter, with the induction hypothesis used in
orders
lower than $n$:
\begin{equation}
\frac{d}{dt}t^3\frac{d^2\alpha_n}{dt^2}\simeq\frac{\kappa \rho ^2\mu ^2}{8}
\frac{d\alpha_{n-2}}{dt}
\label{alfan}
\end{equation}
correct to order $t^{1-\frac{n}{2}}$. This relation proves
the estimate for $\alpha _n$ in (\ref{indu}).
After summation over $n$ this corresponds exactly to
$ \left(\partial^2-\frac{\kappa\mu ^2}{2}\right)\partial _\nu\alpha=0$
which is solved by
(\ref{an1}). As seen in \ref{higher} this does not guarantee exponential
fall off; also exponential rise is possible, unless a particular solution
is picked in low orders. However, the requirements necessary for obtaining
the desired asymptotic behavior of the full solution have been met
in the present case by the choice of integration constants in Sec.
\ref{iteration}
and by modifying the second-order equation according to (\ref{alfa2mod}).

For $n=2$ and $n=4$ the leading terms are of order $t^0$ and $t^{-1}$,
respectively. In each of these cases (\ref{alfan}) only
contains a nonlogarithmic term, and the equation only restricts the
logarithmic
parts of the leading terms; the nonlogarithmic parts have to be
fixed by
adjustment of integration constants, as we saw in Sec. \ref{iteration}.
However, for $n>4$ the situation is different. Here the left-hand side
of (\ref{alfan}) is of order $t^{1-\frac{n}{2}}$ and contains in each
case both a logarithmic
and a nonlogarithmic term, and both the logarithmic and the nonlogarithmic part
of $\alpha _n$ is determined by the equation. Consequently, no further
adjustment is
necessary to produce (\ref{an1}) as the only solution of (\ref{alfan})
after summation over $n$, and no restrictions on the integration constants
$c_{n;1}$ occur at these orders.

As an example we will use (\ref{alfan})
 to determine the leading terms of $\alpha_6$.
Keeping only the leading terms in $\alpha_4$
(which are of order $t^{-1}$) as given in (\ref{CC}) we find the equation
\begin{equation}
\frac{d^2\alpha_6}{dt^2}
  \simeq\frac{\kappa^3\rho^6\mu^6}{1024t^4}
  \left(\log\frac{\kappa\rho^2\mu^2}{8t}+2\gamma-\frac52\right).
\end{equation}
This is integrated
to give the result
\begin{equation}
\alpha_6\simeq\frac{\kappa^3\rho^6\mu^6}{6144t^2}
  \left(\log\frac{\kappa\rho^2\mu^2}{8t}+2\gamma-\frac{10}{3} \right).
\end{equation}
Comparing this to (\ref{sumk}) we find that $\alpha_6$ is
indeed the sixth order term in
$\rho^2\frac{m_{vec}}{x}K_1(m_{vec}x)$.
Note that this result is obtained without tuning the integration
constants.

Next  the same analysis is applied to (\ref{li2}).
The equation obeyed by the leading terms is to order $n$:
\begin{equation}
\frac{d^2f_n}{dt^2}\simeq\frac{\rho ^2\mu ^2}{4t^3}f_{n-2}
\label{FN}
\end{equation}
correct to order $t^{-\frac{1}{2}-\frac{n}{2}}$, which matches
(\ref{an2}) with (\ref{li2}), and the choice of integration
constants in low orders fixes the asymptotic behavior according to
(\ref{an2}).

%%%%%%%%%%%%%%%%%%%%%%%%%%%%%%%%%%%%%%%%%%%
\subsubsection{$t\rightarrow \infty$} \label{infty}
%%%%%%%%%%%%%%%%%%%%%%%%%%%%%%%%%%%%%%%%%%%%%

In this limit it is checked  that $\alpha _n,\hspace{1 mm} n \neq 0$,
and
$\sqrt{t}f_ n$
diverge at most logarithmically. This has already been proven in
Sec. \ref{iteration} in the cases $n=2,3,4$. For $n>4$ the statement is
demonstrated by induction.

For $n>4$ the induction hypothesis in combination with (\ref{li1}) leads to
\begin{eqnarray}
  &&\frac{d}{dt}\left(\left(\frac{t}{1+t}\right)^3
\left(\frac{d^2\alpha_n}{dt^2}-\frac{3\alpha _2}{1+t}
\frac{d^2\alpha_{n-2}}{dt^2}-\frac{3\alpha _4}{1+t}
\frac{d^2\alpha_{n-4}}{dt^2}\right .\right .
\nonumber \\ && \left .\left .+\frac{6\alpha _2^2}{(1+t)^2}
\frac{d^2\alpha_{n-4}}{dt^2}+\cdots\right)\right)
  =O(t^{-4}).
\end{eqnarray}
The induction hypothesis implies that all terms on the left-hand side
excluding the first one are at most $O(t^{-4})$.
Thus, rewriting this equation after the first integration:
\begin{eqnarray}
 \frac{d^2\alpha_n}{dt^2}\simeq O(t^{-3})+c_{n;1}
 \left(\frac{1+t}{t}\right)^3
\end{eqnarray}
one sees that the terms $c_{n;1}(\frac{3}{t}+1)$ give rise to unwanted terms in
$\alpha _n$ of the form $c_{n;1}(3t\log t+\frac{1}{2}t^2)$. Similar unwanted
terms were discarded from $\alpha _2$ and $\alpha _4$, leading to the
modified equation
(\ref{alphaeq1}) instead of (\ref{li1}) for the determination of $\alpha $,
with the constant
$\bar{\sigma}$ given by (\ref{alphaeq2}). This procedure can also be
applied here, with
the same conclusion. The argument shows that the $O(\mu ^6)$
terms of $\bar{\sigma}$ are arbitrary since they are given by
the arbitrary constants $c_{n;1}$.

Similarly, upon examining equation (\ref{li2}) at order $n$ one obtains
from the induction hypothesis that keeping only the leading terms we can
disregard all terms
of order $t^{-\frac{3}{2}}$ and thus we are left with
\begin{equation}
(1+t)^2\frac{d^2f_n}{dt^2}-\frac{3}{4}f_n=O(t^{-\frac{3}{2}}) \label{flign2}
\end{equation}
whence
\begin{equation}
\frac{f_n}{(1+t)^\frac{3}{2}}=
O(t^{-2}).
\end{equation}
Thus, $\sqrt{t}f_n$ can diverge at most  logarithmically, and the proof
by induction is completed.

%%%%%%%%%%%%%%%%%%%%%%%%%%%%%%%%%%%%%%
\subsection{Modified equations}\label{modeq}
%%%%%%%%%%%%%%%%%%%%%%%%%%%%%%%%%%%%%%

In Secs. \ref{iteration} and \ref{limit} it was found that in order
to obtain
a finite action solution of eqs. (\ref{li1}) and (\ref{li2}) that reduces
to the usual
instanton in the massless limit,  one must modify the field equation
(\ref{li1})  to (\ref{alphaeq1}),
while (\ref{li2}) requires no
modification. In the light of this, it should be investigated which types
of constraints can lead to this modification.

Several gauge-invariant constraints have been considered in
the litterature. Two of these are examined and shown to lead to an infinite
action.
The Yang-Mills field equation with the general expression for the constraint
modification $\sigma \int d^4xO$ included in the action is:
\begin{eqnarray}
&&-\bar{\eta}^{a}_{\mu\nu}\alpha^2\partial_\nu\left(\alpha^{-3}\partial^2\alpha
  \right)+\bar{\sigma }\frac{\delta }{\delta A_\mu^a}\int d^4xO
  \nonumber \\&&
  =-\frac{\kappa}{2}f^2\bar{\eta}^{a}_{\mu\nu}\partial_\nu\log\alpha
\label{fe}
\end{eqnarray}
with $\bar{\sigma }$ a suitable constant.

First the constraint used by Klinkhamer \cite {Klinkhamer} is considered:
\begin{equation}
O_K=\left(\frac18 \epsilon_{\mu\nu\kappa\tau}
G^a_{\mu\nu}G^a_{\kappa\tau}     \right)^2
\end{equation}
 To zeroth order in the mass one finds:
\begin{equation}
\frac{\delta }{\delta A_\mu^a}\int d^4xO_K=-\frac{3072}{\rho^8}\bar
{\eta}^{a}_{\mu\nu}
  x_\nu\left (\frac{t}{1+t}\right )^7
\end{equation}
Inserting this into the field equation (\ref{fe}) gives the equation for
$\alpha_2$ where we have substituted $\alpha^2\rightarrow\alpha_0^2$ in the
constraint term:
\begin{eqnarray}
\frac{d}{dt}\left(\left(\frac{t}{1+t}\right)^3\frac{d^2\alpha_2}{dt^2}\right)
  +\frac{\bar{\sigma }_Kt^5}{(1+t)^9}
=\frac{\kappa\rho^2\mu^2}{8(1+t)^4}. \label{Klinkhamer}
\end{eqnarray}
with $\bar{\sigma }_K$ a constant of order $\mu ^2$. It is impossible, with
this equation
determining
$\alpha _2$, to eliminate both the term $\frac{1}{2}c_{2;1}t^2$ and
$3c_{2;1}t\log t$ in (\ref{alfatofoerst}), which is necessary for a
finite action solution, and still have Bessel function behavior at
infinity, so this constraint must be rejected.

Similarly, the constraint used by Aoyama et al. \cite{Aoyama} is considered:
\begin{equation}
O_A=\epsilon^{abc}G^a_{\mu\nu}G^b_{\nu\rho}G^c_{\rho\mu}
\end{equation}
whence:
\begin{equation}
\frac{\delta }{\delta A_\mu^a}\int
d^4xO_A=\frac{768}{\rho^6}\bar{\eta}^{a}_{\mu\nu}
  x_\nu\left(\frac{t}{1+t}\right)^5
\end{equation}
leading to an equation
\begin{eqnarray}
\frac{d}{dt}\left(\left(\frac{t}{1+t}\right)^3\frac{d^2\alpha_2}{dt^2}\right)
  +\frac{\bar{\sigma }_At^3}{(1+t)^7}
=\frac{\kappa\rho^2\mu^2}{8(1+t)^4}\label{Aoyama}
\end{eqnarray}
that again leads to an infinite action.

The question is now
whether it is possible at all to obtain the modified equations
 from an action
principle with a Lagrangian density expressed only in terms of the field
variables $A^{a}_{\mu}$ and $\phi $. The immediate difficulty here is that
the equation is formulated in terms of the variable $\alpha $ and not the gauge
potential $A^a_\mu $. In order
to handle this  we decompose $A^{a}_{\mu}$ as follows:
\begin{equation}
A^a_\mu=\bar{\eta}^{a}_{\mu \nu }(\alpha_\nu -\partial _\nu \log \alpha )+
{\cal A}_{a\mu }
\label{dekomb}
\end{equation}
with
\begin{equation}
\partial _\nu \alpha _\nu =0,\hspace{0.1 mm}
 {\cal A}_{ai }={\cal A}_{ia },\hspace{0.1 mm}
\delta _{ai}{\cal A}_{ai }=0,\hspace{0.1 mm}i=1,2,3.
\end{equation}
The ansatz (\ref{ansatz}) is regained if one sets $\alpha _\nu =0$ and
${\cal A}_{a\mu }=0$.

The Yang-Mills field equation can be obtained by
insertion of (\ref{dekomb}) into the Lagrangian (\ref{lagr}), if one takes
variation of (\ref{lagr})
with respect to $\log \alpha$ (or $\alpha$), $\alpha_\nu $
and ${\cal A}^{a}_{\mu }$, and next sets $\alpha _\mu=0$ and ${\cal
A}_{a\mu}=0$.
Since the constrained instanton is expressed in terms of $\alpha $,
the field components $\alpha_\nu $
and ${\cal A}_{a\mu }$ act as Lagrange multiplier fields as far as the
formulation of the constraint goes. Thus the difficult part of the construction
of a good constraint concerns the part of the constraint
involving only
 $\alpha$.

 With
$$S_{YM}=\int L_{YM}d^4x,\hspace{0.1 mm}
L_{YM}=-\frac{1}{g^2}\frac{1}{4}G_{\mu\nu}^aG_{\mu\nu}^a $$ one obtains
\begin{equation}
\frac{\delta S_{YM}}{\delta \alpha (x)}
  =-\frac{1}{g^2}\frac{3}{\alpha }
  \partial _\mu
  \left(\alpha ^2\partial _\mu
  \left(\frac{\partial ^2\alpha (x)}{\alpha ^3}\right)\right).
\end{equation}
This should be compared to the modified equation (\ref{alphaeq1}) in terms of
the variable $x$:
\begin{eqnarray}
  &&\partial _\mu\left(\alpha^{-3}
\partial ^2\alpha -
 \frac{\bar{\sigma }}{\rho
^2}\left(\frac{\rho^2}{x^2+\rho^2}\right)^3\left(1+\frac{3x^2}{\rho^2}
  \right)\right)
  \nonumber \\&&  =\frac{\kappa}{2}
  f^2\alpha^{-3}\partial_\mu \alpha .
\label{dom}
\end{eqnarray}
Thus we learn that we want to add a term
$S_{const}$ to
the action (the subscript "const" indicates constraint) such that
\begin{equation}
\frac{\delta S_{const}}{\delta \alpha }=\frac{1}{\alpha }
\partial _\mu \left(\alpha ^2\partial _\mu F(\alpha )\right)
\label{SKonst}
\end{equation}
where to second order in the mass variable
\begin{equation}
F(\alpha )\simeq\frac{3\bar{\sigma
}_2}{g^2\rho^2}\left(\frac{\rho^2}{x^2+\rho^2}
\right)^3
  \left(1+\frac{3x^2}{\rho^2}\right).
\label{Falfa}
\end{equation}
with $\bar{\sigma }_2$ the part of $\bar{\sigma}$ as given in (\ref{alphaeq2})
that is of
second order in $\mu ^2$.

Eq. (\ref{SKonst}) contains two partial derivatives. Consequently
$S_{const}$ can contain the term $\sigma S_a$ with
\begin{equation}
S_{a}=\int g(\alpha )\partial_\mu \alpha \partial_\mu \alpha d^4x
\label{Skonst}
\end{equation}
that is compared with (\ref{SKonst}) in lowest order in the mass variable,
where the procedure
leading to (\ref{Klinkhamer}) and (\ref{Aoyama}) is repeated and where
$\sigma $ is
fixed at the value $\bar{\sigma }_2$.
The two formulas are equivalent for
\begin{equation}
g(\alpha )=-\frac{18}{g^2\rho ^2}(-\frac{2}{3\alpha ^3}+\frac{1}{2\alpha ^2}).
\end{equation}

The  fields $\alpha_\nu $
and ${\cal A}_{a\mu }$ only have to enter the constraint linearly.
This is accomplished if one adds to the action a term $\sigma S_b$ with
\begin{eqnarray}
&&S_{b}=-\int d^4x
(\bar{\eta}^{a}_{\mu \nu }\alpha_\nu +
{\cal A}_{a\mu })\bar{\eta}^{a}_{\mu\lambda}\alpha ^2
\partial _\lambda (1-\frac{1}{\alpha })^3 \times
\nonumber \\&& \times
(1+\frac{3}{\alpha -1}).
\label{Llagra}
\end{eqnarray}
The total constraint action is according to this prescription
\begin{equation}
S_{const}=\sigma (S_a+S_b-c) \label{constig}
\end{equation}
where the Lagrange multiplier $\sigma$ should be fixed to
$\bar{\sigma }$.

The constraint (\ref{constig}) does not have a very convenient form
since it is desirable that the twelve components of the gauge field
enter the constraint on the same footing. We have not succeeded in
constructing a good constraint with this property. Obvious candidates like
(\ref{Klinkhamer}) and (\ref{Aoyama}) fail to produce a finite action.

%%%%%%%%%%%%%%%%%%%%%%%%%%%%%%%%%%%%%%%%%%
\subsection{A constraint that almost works} \label{constr}
%%%%%%%%%%%%%%%%%%%%%%%%%%%%%%%%%%%%%%%%%%
Instead of a  constraint involving the field variables alone it is also
possible,
as pointed out by Wang
\cite {Wang},
to use a source-type constraint, where the whole quantum field
enters the constraint linearly.

In order to obtain a pure source term in the modified Yang-Mills field equation,
one should adjust the extra term in (\ref {alphaeq1}), converting it into:
\begin{eqnarray}
  \alpha ^2\frac{d}{dt}\left(\alpha^{-3}t^3\frac{d^2\alpha}{dt^2}\right)-
\frac{3\bar{\sigma }t}{2(1+t)^2}
  =\frac{\kappa\rho^2}{8}
  f^2\alpha^{-1}\frac{d\alpha}{dt}.  \label{alphaeq10}
\end{eqnarray}
This is accomplished if one adds to the action a term
\begin{equation}
S_{const}=\sigma\left(\Sigma_{\rm prov}[A]-c\right)
  \label{scon}
\end{equation}
where $\Sigma_{\rm prov}[A]$ is defined by
\begin{equation}
\Sigma_{\rm prov}[A]=-12\int d^4x A^a_{\nu}\bar{\eta}^{a}_{\nu\lambda}
  \frac{\rho^2x_\lambda}{x^2(x^2+\rho^2)^2}
\label{Sigma}
\end{equation}
and where the Lagrange multiplier $\sigma$ should be fixed to
$\bar{\sigma }$ (this provisional constraint will be completed in
Sec. \ref{lesub}).
Comparing $\Sigma_{\rm prov}$ to the massless instanton solution
(\ref{masslessinstanton})
one finds
\begin{equation}
\Sigma_{\rm prov}[A]=\frac{3}{\rho }\int d^4xA^{a}_{\nu }(x)
(\frac{\partial A^a_{0\nu
}(x)}{\partial \rho }
-\frac{2}{\rho }A^a_{0\nu }(x)).
\label{SSSigma}
\end{equation}

%%%%%%%%%%%%%%%%%%%%%%%%%%%%%%%%%%%%%%%%%%%%%%%
\subsubsection{Modified limit considerations}
%%%%%%%%%%%%%%%%%%%%%%%%%%%%%%%%%%%%%%%%%%%%%%%
Since we have swapped $\alpha _0$ for $\alpha $ in (\ref{alphaeq10}) compared to
 (\ref {alphaeq1}) we have to check that the limit considerations of Sec.
\ref{limit} still hold true.
Orders zero and two are exactly as in Sec. \ref{iteration} but something
new appears in the fourth order equation:
\begin{eqnarray}
&&\frac{d}{dt}\left(\left(\frac{t}{1+t}\right )^3\frac{d^2\alpha_4}{dt^2}
  -3\alpha_2\frac{t^3}{(1+t)^4}\frac{d^2\alpha_2}{dt^2}\right)
  \nonumber \\&&
  -\frac{3\bar{\sigma }_4t}{2(1+t)^4}
  +\alpha_2\frac{3\bar{\sigma }_2t}{(1+t)^5}
\nonumber \\
 && \;\;=\frac{\kappa\rho ^2\mu ^2}{8}\left(\frac{d}{dt}
    \left(\frac{\alpha_2}{(1+t)^4}\right)+\frac{\alpha_2}{(1+t)^5}
    +\frac{2f_3}{\mu (1+t)^\frac{7}{2}}\right).
    \label{146}
\end{eqnarray}
The $\bar{\sigma }_2$-term is new, so we must verify that this new term does not
ruin the
integrability at the origin or the Bessel function behavior at infinity.
In the limit $t\rightarrow 0$ the new term goes as $t\log t$ which is
subleading. In the other limit, $t\rightarrow \infty$, this term
vanishes as $t^{-3}$ so it can only give allowed logarithmic contributions to
$\alpha_4$.
This analysis shows again that the constrained instanton solution is uniquely
determined up to $O(\mu ^3)$ but ambiguous in higher orders.
To a general order $n>4$ one can perform the analysis carried out in
connection with
(\ref{146}) with the same conclusion.

Thus, by adding to the action (\ref{scon}) one
finds a solution to the modified field equations with a finite action,
and
which gives the massless instanton in the $\mu\rightarrow 0$ limit.

%%%%%%%%%%%%%%%%%%%%%%%%%%%%%%%%%%%%%%%%%%%
\subsection{Leading vs. subleading terms} \label{lesub}
%%%%%%%%%%%%%%%%%%%%%%%%%%%%%%%%%%%%%%%%%%
Until now the exponential fall off at infinity valid to all orders in the
mass only takes into account the leading terms to each order.
An analysis similar to that of
(\ref{leadvssub}) is carried out  in this section showing that when the sum
of the
leading terms vanishes exponentially in
the limit $x\rightarrow\infty$, then so
does the sum of the subleading terms, provided that the constraint is
modified such
that the constraint term of the field equation explicitly shows exponential
fall off.

The final form of the Yang-Mills field
equation  (\ref{alphaeq})
including the extra term from Sec. \ref{constr} multiplied with
an exponential
factor is taken as:
\begin{eqnarray}
&&\alpha\partial_\nu\partial^2\alpha
  -3\partial_\nu\alpha\partial^2\alpha-
  \frac{\kappa }{2}f^2\alpha\partial_\nu\alpha
   =\alpha ^2S_{\nu}. \label{finaleq}
\end{eqnarray}
where we define the source $S_{\nu}$:
\begin{equation}
S_\nu =-\frac{12\bar{\sigma }\rho ^2x_\nu }{x^2(x^2+\rho ^2)^2}
e^{-km_{vec}^2x^2} \label{S}
\end{equation}
with $k$ an arbitrary real positive number. The Higgs field
equation (\ref{feq})
is unchanged. This gives the following equations
for the
nextleading terms
\begin{eqnarray}
&&\partial_\nu(\partial^2-\frac{\kappa \mu^2}{2})\alpha^{(4)}=
  3\partial_\nu\alpha^{(2)}\partial^2\alpha^{(2)}
  +\kappa \mu f^{(2)}\partial_\nu\alpha^{(2)}
  \nonumber \\ &&
  -\frac{12\bar{\sigma }_2\rho^2x_\nu}{x^6}
  e^{-km_{vec}^2x^2}.
\end{eqnarray}
\begin{eqnarray}
&&(\partial^2-\mu^2)f^{(4)}
  =-2\alpha^{(2)}\partial^2f^{(2)}+\frac{3}{4}\mu(\partial_\nu\alpha^{(2)})^2
  \nonumber \\ &&
+\frac{1}{2}\left(4\mu^2\alpha^{(2)}f^{(2)}+3\mu(f^{(2)})^2\right).
\end{eqnarray}
This modification follows if (\ref{scon}) is added to the action in a form where
the integrand in (\ref{Sigma}) is multiplied by the exponential factor
$e^{-km_{vec}^2x^2}$. The final form of the constraint functional is thus
\begin{equation}
\Sigma[A]=-12\int d^4x A_{a\nu}\bar{\eta}_{a\nu\lambda}
  \frac{\rho^2x_\lambda}{x^2(x^2+\rho^2)^2}e^{-km_{vec}^2x^2}.
\label{finSigma}
\end{equation}

As was the case in Sec. \ref{constr} one again has to check that the limit
considerations of Sec. \ref{limit}  still are valid.
In the limit $x\rightarrow \infty$ there is exponential fall off.
Indeed, this was the reason why the exponential factor was inserted in the first
place. In the other limit, $x\rightarrow 0$, the extra exponential factor
becomes unity and thus this constraint also ensures a finite
action.

$\Sigma[A]$ can be calculated order by order
as the equation for $\alpha $ is solved.  In zeroth
order:
\begin{equation}
\Sigma _0=-36\pi^2
\end{equation}
while to second order
\begin{eqnarray}
\Sigma _2=12\pi^2\frac{\kappa \rho ^2\mu ^2}{8}
  \left(\log\frac{\kappa \rho^2 \mu ^2}{8}+2\gamma+2+12k\right).
\end{eqnarray}
Picking a value of the constant $c$ obviously fixes the scale
parameter $\rho $ according to:
\begin{eqnarray}
&&-36\pi^2+12\pi^2\frac{\kappa\rho ^2\mu ^2}{8}
  \left(\log\frac{\kappa\rho ^2\mu ^2}{8}+2\gamma+2+12k\right)
  \nonumber \\ &&
  +O( \rho ^4\mu ^4)=c.
\end{eqnarray}
Taking here
\begin{equation}
c=-36\pi ^2(1+\epsilon)
\end{equation}
with $0<\epsilon <<1$ one obtains a transcendental equation
the solution of which expresses $\rho $ in terms of $\epsilon $
in such a way that $\rho \mu <<1$. In this way the choice of the
constant $c$ fixes the scale.

%%%%%%%%%%%%%%%%%%%%%%%%%%%%%%%%%%%%%%%%%%%%%%%%%%%%%
\subsubsection{Finding the modified $\alpha_4$}
%%%%%%%%%%%%%%%%%%%%%%%%%%%%%%%%%%%%%%%%%%%%%%%%%%%%%%
For the determination of $\alpha_4$ from
(\ref{finaleq}) one solves
 the following equation replacing (\ref{alpha4}):
\begin{eqnarray}
&&\frac{d}{dt}\left(\left(\frac{t}{1+t}\right)^3\frac{d^2\alpha_4}{dt^2}
  -3\alpha_2\frac{t^3}{(1+t)^4}\frac{d^2\alpha_2}{dt^2}\right)
  -\bar{\sigma }_4\frac{3t}{2(1+t)^4}
  \nonumber \\ &&+\bar{\sigma }_2\frac{3t}{(1+t)^5}\alpha_2
   +\bar{\sigma }_2k\frac{\kappa\rho^2\mu^2}{4}\frac{3}{(1+t)^4}
   \nonumber \\ &&=\frac{\kappa\rho ^2\mu ^2}{8}\left(\frac{d\alpha_2}{dt}
    \frac{1}{(1+t)^4}-\frac{3\alpha_2}{(1+t)^5}
    +\frac{2f_3}{\mu (1+t)^\frac{7}{2}}\right),
\end{eqnarray}
with $\bar{\sigma }_4$ a suitable constant. $\alpha _4$ is split according to
\begin{equation}
\alpha_4=\hat{\alpha}_4 +\tilde{\alpha}_4
\label{split}
\end{equation}
with $\hat{\alpha}_4 $ given in (\ref{CC})
while $\tilde{\alpha}_4 $ is
\begin{eqnarray}
&&\tilde{\alpha}_4=\frac{\kappa ^2\rho ^4\mu ^4}{64}
  \left(\log\frac{\kappa\rho ^2\mu ^2}{8}+2\gamma-1\right) \times
  \nonumber \\ && \times
  \left(-\frac{1}{6t}-(1+t)\log\frac{1+t}{t}\right)
  -\frac{(16k+1)\kappa ^2\rho ^4\mu ^4}{768t}
  \nonumber \\
  &&-\frac{\kappa ^2\rho ^4\mu ^4}{64}\left((1+t)
  (\frac32-\log t)\log\frac{1+t}{t}+\log t\right .
  \nonumber \\&& \left .
   +\frac16t
  -\frac16t^2\log\frac{1+t}{t}
  -\frac{\log(1+t)}{6t}-2\Phi\left(\frac{1}{t}\right)\right)
  \label{hathat}
\end{eqnarray}
which has the following asymptotic behavior at $t\rightarrow 0 $:
\begin{eqnarray}
&&\tilde{\alpha}_4\simeq -\frac{\kappa ^2\rho ^4\mu ^4}{384t}
  \left(\log\frac{\kappa\rho ^2\mu ^2}{8}+2\gamma-1\right)
  \nonumber \\ &&
  -\frac{(16k+1)\kappa ^2\rho ^4\mu ^4}{768t}
\end{eqnarray}
that one adds to (\ref {CC}) in order to obtain
the asymptotic behavior of the modified $\alpha _4$.
The outcome should match (\ref{besselfire}).
The constant $c_{4;1} $ must consequently have an additional term
\begin{eqnarray}
&&\frac{\kappa ^2\rho ^4\mu ^4}{192}
  \left(\log\frac{\kappa\rho ^2\mu ^2}{8}+2\gamma+\frac{16k-1}{2}\right)
  \label{added}
\end{eqnarray}
to be  added to (\ref{cIVI}). The resulting value of $c_{4;1}$, and
consequently
of the factor $\bar{\sigma }_4$ in front of the fourth-order constraint,
vanishes
for $k$ chosen according to
\begin{eqnarray}
&&\frac{\kappa^2\rho ^4\mu ^4}{24}k=\frac{\kappa\rho ^4\mu ^4}{192}
(3-\kappa )-\frac{\kappa^2\rho ^4\mu ^4}{96}\log \frac{\kappa }{2}
\nonumber\\&&
-\frac{\kappa\rho ^4\mu ^4}{96}(\kappa -1)\left(\log\frac{\rho ^2\mu ^2}{4}
+2\gamma \right )
\end{eqnarray}
where each term on the right-hand side is positive for $1<\kappa <2$.

With the calculation in this subsection it is clear
that different constraints producing a finite action solution have different
subleading terms. In fact, it is possible to remove the subleading terms from
$\alpha_4$
if the exponential of (\ref{finaleq}) also contains a term
$$-k_1\frac{\kappa\mu^2\rho^2}{2}\frac{\log t}{t}$$
in the exponent, with $k_1$ a positive constant.

%%%%%%%%%%%%%%%%%%%%%%%%%%%%%%%%%%%%%
\subsubsection{An alternative approach}
%%%%%%%%%%%%%%%%%%%%%%%%%%%%%%%%%%%%%
Writing the constraint term of (\ref{li1}) as is suggested in (\ref{Llagra})
\begin{equation}
-\frac{\sigma}{4}(1-\frac{1}{\alpha})^3(1+\frac{3}{\alpha-1})   \label{enmod}
\end{equation}
one can check that the subleading terms vanish exponentially.
This form of the constraint can also be obtained through a source term.
One starts with a modifiation term
\begin{equation}
-\frac{\sigma}{4}(1-\frac{1}{\alpha_0})^3(1+\frac{3}{\alpha_0-1})
\end{equation}
and uses this to obtain $\alpha_2$. Then one modifies the constraint to get
a modification term
\begin{equation}
-\frac{\sigma}{4}(1-\frac{1}{\alpha_0+\alpha_2})^3(1+\frac{3}
{\alpha_0+\alpha_2-1})
\end{equation}
and uses this to find $\alpha_4$. Then one again modifies the constraint to
find $\alpha_6$, $\alpha_8$ etc. In this way we can effectively work with the
modification term (\ref{enmod}).

The integration constant $c_{4;1}$ and thus the fourth-order constraint are
with this approach modified compared to (\ref{cIVI}) with (\ref{added})
added; also
the subleading terms contain additional terms to those found
in the previous section. This emphasizes the ambiguity of the constraint
beyond lowest order.

%%%%%%%%%%%%%%%%%%%%%%%%%%%%%%%%%%%%%%%%%%%%%%%%%%%%%
\subsubsection{Effective Lagrangian}
%%%%%%%%%%%%%%%%%%%%%%%%%%%%%%%%%%%%%%%%%%%%%%%%%%%%%%

A systematic way of representing subleading terms at large distances is like
in the scalar case obtained by iteration of field equations. It is here
convenient
to use a field $a_\nu= \partial _\nu \log \alpha $, in terms of which
the field equations are
\begin{equation}
\partial ^2 a_\nu-2a_\nu \partial _\lambda a_\lambda-\partial_\nu (a_\mu a_\mu )
-2a_\nu a_\mu a_\mu-\frac{\kappa }{2}f^2a_\nu=S_\nu,
\end{equation}
\begin{equation}
\partial ^2 f-\frac{3}{4}a_\mu a_\mu f+\frac{1}{2}(\mu ^2-f^2)f=0,
\end{equation}
with $S_\nu $ defined in (\ref{S}).
These field equations are obtained from an effective Lagrangian
\begin{eqnarray}
&&L_{eff}=-\frac{1}{2}\partial _\mu a_\nu\partial _\mu a_\nu
-\partial_\nu a_ \nu a_ \mu a_ \mu -\frac{1}{2}a_\nu a_\nu a_\mu a_\mu
\nonumber \\ &&-\frac{\kappa }{4}f^2a_\nu a_\nu-\frac{2\kappa }{3}
\partial _\mu f\partial _\mu f-\frac{\kappa }6(f^2-\mu ^2)^2+a_\nu S_\nu .
\end{eqnarray}

%%%%%%%%%%%%%%%%%%%%%%%%%%%%%%%%%%%%%
\subsection{The 't Hooft path integral measure} \label{thooft}
%%%%%%%%%%%%%%%%%%%%%%%%%%%%%%%%%%%%%
For completeness we briefly indicate how the 't Hooft \cite {'t Hooft} path
integral measure is obtained from our analysis. This subsection
mostly contains well-known results. However, we indicate how our methods
can be used to generalize the result of 't Hooft.

The value of the classical action up to second order in the mass
parameter when calculated by means of the constrained instanton solution of
Sec.
\ref {iteration} leads to the result
\begin{equation}
S=-\frac{8\pi^2}{g^2}-\frac{2\pi^2\kappa\rho^2\mu^2}{g^2}
\end{equation}
in agreement with \cite {'t Hooft}, \cite {Affleck}, \cite {Espinosa}.
Beyond second order the result will contain ambiguities since it depends on
the form
of $\alpha _4$ which, as we have seen, can be modified.

The Euclidean path integral
\begin{equation}
Z=\int [dA][d\phi]e^{-S_E[A,\phi]}
\end{equation}
with $S_E[A,\phi]$ the Euclidean Yang-Mills-Higgs action is
evaluated by the saddle-point method.
For this purpose the previously determined constrained instanton
solution is used
 by a Faddeev-Popov trick.
We  write unity as
\begin{equation}
1=\int d\rho\Delta[A,\rho]\delta\left(\Sigma [A]-c\right)
  =\int d\rho\Delta[A,\rho]\delta\left(\Sigma [A-\bar{A}]\right)
\end{equation}
where the constrained instanton is denoted $(\bar{A},\bar{\phi})$, and where the
constraint it obeys  was used in the last step.

Multiplying the path integral by the Faddeev-Popov unity one obtains
\begin{equation}
Z=\int [dA][d\phi]\int d\rho\Delta[A,\rho]\delta\left(
  \Sigma [A]-c\right)e^{-S_E[A,\phi]}
\end{equation}
where $\Sigma [A]$ was defined in (\ref{finSigma}), while
\begin{equation}
\Delta[A,\rho ]
  =\left| 12\int d^4x A_{a\nu}\bar{\eta}_{a\nu\lambda}
  \frac{\partial }{\partial \rho }\frac{\rho^2x_\lambda}{x^2(x^2+\rho^2)^2}
  \right|
\label{Delta}
\end{equation}
to order $\mu ^0$.
It is known from the previous analysis that the modified
action
\begin{equation}
\tilde{S}_E[A,\phi]=S_E[A,\phi]+\bar{\sigma }\Sigma\;,\;\;
  \bar{\sigma }=-\frac{\kappa\rho^2\mu^2}{6}+\cdots
\end{equation}
has a finite solution, so using the $\delta$-function we write
$S_E=\tilde{S}_E-\bar{\sigma }c$ and the path integral is hence
\begin{equation}
Z=\int [dA][d\phi]\int d\rho
  \Delta[A,\rho]\delta\left(\Sigma [A-\bar{A}]\right)
  e^{-\tilde{S}_E[A,\phi]}
  e^{\bar{\sigma }c}.
\end{equation}
New integration variables are introduced
through the substitution $$(A,\phi)\rightarrow
(\bar{A}+A,\bar{\phi}+\phi).$$ In the Gaussian approximation
\begin{equation}
\Delta[A+\bar{A},\rho]\simeq \Delta[\bar{A},\rho]
\end{equation}
where terms depending on $A$ have been disregarded in $\Delta$,
and
\begin{eqnarray}
&& Z=e^{-S_E[\bar{A},\bar{\phi}]}\int [dA][d\phi]\int d\rho \Delta[\bar{A},\rho]
  \delta(\Sigma [A])\times
  \nonumber \\ && \times
  e^{-(\tilde{S}_E[\bar{A}+A,\bar{\phi}+\phi]-\tilde{S}_E[\bar{A},\bar{\phi}])}.
\label{vej}
\end{eqnarray}

The path integral (\ref{vej}) is the same as that given by 't Hooft
\cite{'t Hooft} in the approximation where the classical action
$S_E[\bar{A},\bar{\phi}]$ is
computed up to order $\mu ^2$ and the fluctuations (including the
Faddeev-Popov determinant
$\Delta [A, \rho ]$) to order $\mu ^0$. In this
approximation
the integral over $\rho $ is identical to what is obtained by means of the
standard method
of collective coordinates. This follows from (\ref{SSSigma}), where the term
$\frac{\partial A_{0a\mu }}{\partial \rho}$ in the integrand projects out the
dilatation zero mode. Also $\Delta[\bar{A},\rho]$ is easily checked to be the
same as the corresponding expression obtained by means of collective
coordinates.
This argument is actually independent of the detailed form of the constraint,
provided the projection of the derivative of the constraint with respect to the
gauge
field onto the dilatation zero mode is nonvanishing.

%%%%%%%%%%%%%%%%%%%%%%%%%%%%%%%%%%%%%%%%%%%
\section{Conclusion }
%%%%%%%%%%%%%%%%%%%%%%%%%%%%%%%%%%%%%%

Our results can be summarized in the following way:

For $\phi ^4$ theory
 a finite action instanton solution of the massive theory
without any constraint does not exist because of its
large-distance behavior that is characterized by exponential increase
instead of fall off. Two types of constraint are considered.
If the constraint is required to depend only on the scalar field,
the only possible way to cure this defect
is by means
of a constraint cubic in the field. Other constraints
only depending on the field
are ruled out
because they lead to singular behavior of the constrained instanton
solution at the origin.
The constraint can also amount to having a source term in the
field equation.
This type of constraint can be constructed in such a way
that it has the same effect
as the constraint cubic in the field referred to above.

For the Yang-Mills-Higgs theory the situation is rather different.
Here we found
that exponential fall off at infinity can be obtained by adjustment of
integration constants without imposition of any constraint. On the
other hand, a constraint is necessary in order to ensure absence of
singularities of the constrained
instanton at small distances that prevent the action from being
finite.
The form of the constraint required for this purpose is uniquely determined
to lowest order in the mass variable, and
only a special constraint involving field variables only can be constructed.
On the other hand, a constraint corresponding to a source term
in the Yang-Mills field equation is possible; the explicit form
of the modified field equation is given in (\ref{finaleq}).
The source term in (\ref{finaleq}) can be modified somewhat, and
the constrained instanton is correspondingly not uniquely determined
in and above fourth order in the mass. No modification of the Higgs
field equation is necessary.

{\bf Acknowledgement.} We are grateful to Professor P. van Nieuwenhuizen
 at the Institute for Theoretical Physics, State University
of New York at Stony Brook, where this work was initiated, for his
hospitality and for very helpful discussions.

%%%%%%%%%%%%%%%%%%%%%%%%%%%%%%%%%%%%%%%%%%%
\appendix
%%%%%%%%%%%%%%%%%%%%%%%%%%%%%%%%%%%%%%

%%%%%%%%%%%%%%%%%%%%%%%%%%%%%%%%%%%%%%%%%%%
\section{Modified Bessel functions} \label{Besselapp}
%%%%%%%%%%%%%%%%%%%%%%%%%%%%%%%%%%%%%%
The modified Bessel equation of order unity\cite{Watson}
\begin{equation}
x^2\frac{d^2}{dx^2}f(x)+x\frac{d}{dx}f-(x^2+1)f=0
\label{Bessel}
\end{equation}
has as linearly independent solutions the two modified Bessel
functions
\begin{equation}
I_1(x)=\frac{x}{2}\sum_{n=0}^{\infty}\frac{1}{n!(n+1)!}
  \left(\frac{x^2}{4}\right)^{n}
\label{sumi}
\end{equation}
and
\begin{eqnarray}
&&K_1(x)=\left(\log(\frac{x}{2})+\gamma\right)I_1(x)+\frac{1}{x}
\nonumber \\ &&
-\frac{x}{4}
\sum _{n=0}^{\infty}\frac{1}{n!(n+1)!}\left(\sum_{k=1}^{n+1}\frac{1}{k}
+\sum_{k=1}^{n}\frac{1}{k}\right)\left(\frac{x^2}{4}\right)^n
\label{sumk}
\end{eqnarray}
where $\gamma $ is Euler's constant and the first term in the sum is
$\frac{x}{4}$.
For $x\rightarrow \infty$:
\begin{equation}
I_1(x)\simeq \sqrt{\frac{1}{2\pi x}}e^{x},
\label{asi}
\end{equation}
\begin{equation}
K_1(x)\simeq \sqrt{\frac{\pi }{x}}e^{-x}.
\label{ask}
\end{equation}

The Klein-Gordon equation in four-dimensional Euclidean space
\begin{equation}
(\partial^2-m^2)\phi=0, \hspace{0.1 mm}x\neq 0. \label{KG}
\end{equation}
is with $\phi $ only a function of $x=|x|$ and
 writing
\begin{equation}
\phi (x)=\frac{m}{x}G(mx)
\end{equation}
converted into
\begin{eqnarray}
&&(\frac{d^2}{dx^2}+\frac{3}{x}\frac{d}{dx}-m^2)\phi (x)
\nonumber \\&&
=\frac{m^4}{\xi ^3}
\left(\xi ^2\frac{d^2G(\xi )}{d\xi ^2}+\xi \frac{dG(\xi )}{d\xi }
-(1+\xi ^2)G(\xi )\right)
\end{eqnarray}
with $\xi =mx.$
Here the expression in brackets is recognized as the defining equation of
the modified Bessel functions of order one
(\ref{Bessel}). Consequently the solution of (\ref{KG})
is a linear combination of
$\frac{m}{x }I_1(mx)$ and
$\frac{m}{x }K_1(mx)$.
\end{document}